\documentclass[letterpaper]{article}
\usepackage{amsmath,amssymb}
\usepackage{algpseudocode}
\usepackage{algorithm,url}
\usepackage{graphicx}
\begin{document}
\title{Performance Analysis of Error Control Coding Techniques for Peak-to-Average Power Ratio Reduction of Multicarrier Signals}

\author{Sanjay Singh \thanks{Sanjay Singh is with the Department of Information \& Communication Technology, Manipal Institute of Technology, Manipal-576104, India,Email: sanjay.singh@manipal.edu}~
                M.Sathish Kumar and H.S Mruthyunjaya \thanks{M.Sathish Kumar and H.S Mruthyunjaya is with Department of Electronics \& Communication Engineering, Manipal Institute of Technology, Manipal-576104, India}}

\maketitle
\begin{abstract}
Increasing demands on high data rate mobile communications services will inevitably drive future broadband mobile communication systems toward achieving data transmission rates in excess of 100 Mbps. One of the promising technologies which can satisfy this demand on high data rate mobile communications services is the Orthogonal Frequency Division Multiplexing (OFDM) transmission technology which falls under the general category of multicarrier modulation systems. OFDM is a spectrally efficient modulation technique that can achieve high speed data transmission over multipath fading channels without the need for powerful equalization techniques. However the price paid for this high spectral efficiency and less intensive equalization is low power efficiency. OFDM signals are very sensitive to non-linear effects due to the high peak-to-average power ratio (PAPR), which leads to the power inefficiency in the RF section of the transmitter. This paper analyzes the relation between aperiodic autocorrelation of OFDM symbols and PAPR. The paper also gives a comparative study of PAPR reduction performance of various channel coding techniques for the OFDM signals. For our study we have considered Hamming codes, cyclic codes, convolution codes, Golay and Reed-Muller codes. The results show that each of the channel coding technique has a different PAPR reduction performance. Coding technique with the highest value of PAPR reduction has been identified along with an illustration on PAPR reduction performances with respect to each code. \\
\end{abstract}

\maketitle
\newtheorem{theorem}{Theorem}
\newtheorem{definition}{Definition}
\newtheorem{example}{Example}

\section{Introduction}
With the progressive development of mobile cellular communication technology, the need for multi use services and broadband convergence of voice, data, videos and automated applications in a single device has become obvious. To support these applications there is increasing demand for high bit rate and reliable wireless communication system which has led to many new emerging modulation techniques. One of such techniques is Orthogonal Frequency Division Multiplexing (OFDM). OFDM has emerged as an efficient multicarrier modulation scheme for wireless, frequency selective communication channels. OFDM is a method of transmitting data simultaneously over multiple, equally spaced carrier frequencies using the Fourier transform processing for modulation and demodulation \cite{1}. This method has been proposed or adopted for various types of radio systems such as DAB,DVB \cite{2}\cite{3} and WLAN \cite{4}. Specifically, the OFDM is preferred in most of the high bandwidth, spectrally efficient transmission system due to its robustness in multipath fading environments and hence its effective resistance to inter symbol interference (ISI). It is worth to note that ISI is the single most important factor which limits the maximum bit rate of wireless transmission system, especially in multipath environments.
\par
While OFDM is a promising technology, the principal difficulty encountered in its practical realizations is that, it exhibits a very high peak-to-average power ratio (PAPR). This is due to the summation of the subcarriers, which could get added constructively or destructively depending on the data being transmitted in parallel. This creates large variations between the average and the peak signal power in OFDM systems. Due to this high PAPR, RF power amplifier should be operated in a very large linear region for if not, the signal peak might get into the non-linear region of the power amplifier causing undesirable signal distortion. The operation of the power amplifiers with a very large linear region leads to expensive transmitter and receiver systems. The signal distortion introduced can result in inter modulation noise among the subcarriers and out of band radiation\cite{5}. Furthermore, if battery power is a constraint, as it is the case with portable equipment, then the power amplifier is required to behave linearly up to peak envelope power and hence must be operated inefficiently. Digital hard limiting of the transmitted signal has been shown to alleviate this problem\cite{6} but only at the cost of spectral side lobe growth and consequent performance degradation.  
\par
This leads to the motivation to explore and find other ways to control PAPR of the transmitted signal without compromising the performance of the OFDM systems. A promising method, which has attracted considerable attention, is the use of Error Control Coding (ECC) techniques. The technique of using ECC to counter PAPR, introduced first in\cite{7} and later developed in \cite{8} uses block coding to transmit data across the carriers in such a way that only those polyphase sequences which have low PAPR are selected.Though many ECC based methods \cite{9}\cite{10}\cite{11} have been proposed to date in the literature for PAPR reduction, until now there has been no comparative study of PAPR reduction capability of various channel coding techniques which is a major gap in the research in this area.The objective of this paper is to fill this gap and investigate the PAPR reduction performance for OFDM signal using various channel coding techniques.
\par
The organization of this paper is as follows. In section 2,the model for OFDM signal and its PAPR is discussed. In section 3, PAPR analysis of BPSK modulated OFDM signal is discussed. Section 4 gives notation which has been used throughout section 5. In section 5, various types of linear block codes has been discussed. In section 6, simulation results and its discussion is given. Finally section 7 concludes the paper.

\section {OFDM and PAPR}
For our analysis emphasis is on examining the PAPR of an OFDM signal. Therefore, the OFDM system model used in this paper is a simplified version of the practical OFDM model. Specifically, we have ignored the guard interval because it does not contribute to the PAPR \cite{12}. Assuming that any pulse shaping in the transmitter is flat over all of the subcarriers, and deal only with the PAPR of the baseband signal. 
For one OFDM symbol with N subcarriers, the normalized complex baseband signal can be written as:
\begin{equation}
s(t)=\frac{1}{\sqrt{N}}\sum_{k=0}^{N-1}c_{k}e^\frac{j2\pi kt}{T}    \qquad 0\leq t \leq T
\end{equation}
where \(c_k\) is the frequency domain information symbol mapped to the \(k_{th}\) subcarrier of the OFDM symbol and \(T\) is the OFDM symbol duration. The peak-to-average power ratio (PAPR) of the given frequency domain samples,\\      
$c=\{c_0,c_1,c_2,\cdots, c_{N-1}\}$ is defined as:
\begin{equation}
PAPR\triangleq \stackrel{max}{0\leq t\leq T}\frac{\left|s(t)\right|^2}{E[\left|s(t)\right|^2]}
\end{equation}
where \(E[.]\) denotes a time averaging operator.
The distribution of PAPR values is described using the complementary cumulative distribution function (CCDF). The CCDF of the PAPR represents the probability that the PAPR of a data block exceed a given threshold, $\xi$ and is given \cite{13} by
\begin{equation} 
Pr(PAPR>\xi)=1-(1-e^{-\xi})^N.
\end{equation}

\section{PAPR Analysis of BPSK Modulated OFDM}
For the sake of simplicity we have considered BPSK modulation. PAPR analysis of BPSK modulated OFDM signal is done to understand the reason behind high PAPR. For BPSK modulated OFDM signal, $c_k\in\{-1,+1\}$.
Using the technique as described in \cite{14} and assuming \(T=1.0\) equation (1) can be written as:
{\setlength\arraycolsep{2pt}
\begin{eqnarray}
\sqrt{N}s(t)&=& \sum_{k=0}^{N-1}c_{k}e^{j2\pi kt} \nonumber \\
N\left|s(t)\right|^2&=&\Big(\sum_{k=0}^{N-1}c_k e^{j2\pi kt}\Big)^2 \nonumber \\
&=& \Big(\Re \big[\sum_{k=0}^{N-1}c_k e^{j2\pi kt}\big]\Big)^2+\Big(\Im \big[\sum_{k=0}^{N-1}c_k e^{j2\pi kt}\big]\Big)^2 \nonumber \\
&=& \Big(\sum_{k=0}^{N-1}c_k \cos(2\pi kt)\Big)^2+\Big(\sum_{k=0}^{N-1}c_k \sin(2\pi kt)\Big)^2 \nonumber \\
&=& \sum_{k=0}^{N-1}c_k^2 \cos^2(2\pi kt) + {} \nonumber \\
 && {}+ 2\sum_{k=0}^{N-2}\sum_{i=k+1}^{N-1}c_k c_i \cos(2\pi kt)\cos(2\pi it)+ {} \nonumber \\
&& {} + \sum_{K=0}^{N-1}c_k^2\sin^2(2\pi kt) + {} \nonumber \\
&& {} +2\sum_{k=0}^{N-2}\sum_{i=k+1}^{N-1}c_k c_i \sin(2\pi kt)\sin(2\pi it) \nonumber \\
&=& N+2\sum_{k=0}^{N-2}\sum_{i=k+1}^{N-1}c_k c_i \cos(2\pi (i-k)t) \nonumber \\
&=& N+2P_0(t)
\end{eqnarray}}
where \(\Re[x]\) and \(\Im[x]\) are the real and imaginary parts of \(x\) respectively and the AC component of the power envelope of the OFDM signal $P_o(t)$ is defined as:

\begin{equation}
P_0(t)=\sum_{k=0}^{N-2}\sum_{i=k+1}^{N-1}c_k c_i \cos(2\pi(i-k)t).
\end{equation}
The double summation in (5) can be replaced with a single summation by combining each term to its harmonic and \(P_0(t)\) becomes:
\begin{equation}
P_0(t)=\sum_{k=1}^{N-1}C_k \cos(2\pi kt)
\end{equation}
which can be physically interpreted as the sum of cosine harmonics weighted by the aperiodic autocorrelation \(C_k\) of the frequency domain information bits, where the aperiodic autocorrelation \(C_k\) is defined as:
\begin{equation}
C_k=\sum_{i=0}^{N-k-1}c_i c_{i+k}.
\end{equation}
Substituting for \(P_0(t)\) from (6) in (4), the average power of \(s(t)\) becomes:
\begin{eqnarray} 
E[\left|s(t)\right|^2]&=& E\big[1+\frac{2P_0(t)}{N}\big] \nonumber \\
											&=& 1+\frac{1}{N}E[2P_0(t)] \nonumber \\
											&=& 1.
\end{eqnarray}
As the average power of $s(t)$ is unity, the PAPR in (2), when considering its symmetry with respect to the half symbol time becomes:
\begin{equation}
PAPR=\stackrel{max}{0\leq t \leq 0.5}\big(1+\frac{2}{N}P_0(t)\big)
\end{equation}
where $P_0(t)$ is given by (6). From (6), (7) and (9), it is found that the PAPR is completely characterized by the aperiodic autocorrelations $C_k$. Without any loss of generality we can extend this analysis to other efficient modulation techniques such as QPSK/QAM.
 
For the $N=4$ the aperiodic autocorrelation $C_k$ are given by:
\begin{eqnarray}
C_1&=& c_0 c_1+c_1 c_2+c_2 c_3 \nonumber \\
C_2 &=& c_0 c_2+c_1 c_3	\nonumber \\
C_3&=& c_0 c_3.	\nonumber
\end{eqnarray}
Now there is a need to reduce the aperiodic correlation among the subcarriers so that PAPR is reduced. There are various method to do this such as scrambling \cite{15}, here we have focused only on the linear error control coding techniques. By employing error control coding techniques it gives dual advantage of error control as well as PAPR reduction. Next sections briefs about various error control codes considered for PAPR reduction.
\section{Notation}
We have considered codes over some alphabet $\mathcal{A}$ and reserved the letter $q$ to represent the cardinality of $\mathcal{A}$. In this paper only binary codes have been considered so $q=2$ and $\mathcal{A}=\{0,1\}$. A code $C$ is a subset of $\mathcal{A}^n$ for some positive integer $n$. There are four fundamental parameters associated with a code $C$:
\begin{itemize}
\item Its block length: $n$, where $n\subseteq \mathcal{A}^n$
\item Its message length: $k=\log_q|C|$
\item Its minimum distance: $d_{min}$, which is defined as the the number of coordinates where $c_1$ \& $c_2$ differ, here $c_1,c_2\in C$
\item Its alphabet size,$q=|\mathcal{A}|$
\end{itemize}
A code is often characterized by four parameters it achieves and such code is referred as an $[n,k,d]_q$ code. Bold face letters represents matrices.

\section{Linear Error Control Codes}
There are various ways to specify a code but the most convenient way is to give an encoding function, thereby specifying how to create arbitrary code words. This generalized mapping can be used to describe classes of codes in a succinct manner. In this paper, we have considered a special class of codes called \emph{linear codes}. \\
Liner codes are obtained when the alphabet $\mathcal{A}$ is associated with the field $\mathbb{F}_q$ for some finite value of $q$. In such a case $\mathcal{A}^n$ contains the code words and received words as a vector space.
\begin{definition}
 A subset $L\subseteq \mathbb{F}^n_q$ is a linear subspace of $\mathbb{F}^n_q$ if for every $x,y\in L$ and $\alpha\in \mathbb{F}$ it is the case that $x+y\in L$ and $\alpha.x\in L$.
\end{definition}
When a code $C\subseteq\mathbb{F}^n_q$ is a linear subspace of $\mathbb{F}^n_q$, then the code is called a linear code.
\subsection{Hamming Codes}
For any positive integer $m\geqslant 3$(called parity bits), there exists a Hamming code \cite{16} with the following parameters:
\begin{itemize}
\item Code length: 						$n=2^m-1$
\item Number of message bits: $k=2^m-m-1$
\item Number of parity check-bits:   $m=n-k$
\item Error-correcting capability:   $t=1 (d_{min}=3)$.
\end{itemize}
The parity-check matrix $\textbf{H}$ of this code consist of all the nonzero $m$-tuples as its columns. In systematic form, the columns of $\textbf{H}$ are arranged in the form of $\textbf{H}=[\textbf{I}_m\quad \textbf{Q}]$, where $\textbf{I}_m$ is an $m\times m$ identity matrix, and the submatrix $\textbf{Q}$ consists of $k$ columns that are the $m$-tuples of weight 2 or more. In systematic form, the generator matrix of the code is $\textbf{G}=[\textbf{Q}^T \quad\textbf{I}_k]$. The Hamming code corresponding to the message vector $\textbf{u}$ of $k$ tuple is given by $c=\textbf{u}.\textbf{G}$.
\subsection{Cyclic Codes}
If the components of an $n$-tuple $\textbf{v}=(v_0,v_1,\cdots,v_{n-1})$ is cyclically shifted by one place to the right, the resulting $n$-tuple is $\textbf{v}^{(1)}=(v_{n-1},v_0,v_1,\cdots,v_{n-2})$, which is called a cyclic shift of $\textbf{v}$.
\begin{definition}
An $(n,k)$ linear code $C$ is called a cyclic code if every cyclic shift of a code word in $C$ is also a codeword in $C$. 
\end{definition}
The codeword $\textbf{v}=(v_0,v_1,\cdots,v_{n-1})$ can be written in a polynomial form as:\\ $\textbf{v}(X)=v_0+v_1X+v_2X^2+,\cdots,+v_{n-1}X^{(n-1)}$. 
\begin{theorem}
In an $(n,k)$ cyclic code, there exists one and only one code polynomial of degree $n-k$, $g(X)=1+g_1X+g_2X^2+\cdots+ g_{n-k-1}X^{(n-k-1)}+X^{(n-k)}$.
\end{theorem}
It follows from Theorem (1) that every code polynomial $\textbf{v}(X)$ in an $(n,k)$ cyclic code can be expressed in the form of $\textbf{v}(X)=\textbf{u}(X)g(X)$, where $\textbf{u}(X)$, $u_0,u_1,\cdots,u_{k-1}$ are the $k$ information bits to be encoded, $\textbf{v}(X)$ is the corresponding code polynomial. The polynomial $g(X)$ is called the generator polynomial of the code. The degree of $g(X)$ is equal to the number of parity-check bits of the code. To encode a message polynomial $\textbf{u}(X)$ of $k$-bits into a cyclic code polynomial $\textbf{v}(X)$ of $n$ bits following steps are followed:
\begin{enumerate}
\item Compute $I(X)=X^m\textbf{u}(X)$, to left shift message bits by $k$ positions
\item Compute $\frac{I(X)}{g(X)}$ such that \\$I(X)=q(X)g(X)+r(X)$, where $q(X)$ is quotient polynomial and $r(X)$ is the remainder polynomial
\item Code polynomial $\textbf{v}(X)$ is obtained as $\textbf{v}(X)=I(X)+r(X)$.
\end{enumerate}
 
\subsection{Convolution Codes}
A convolution code is described by three parameters $n,k$ and $K$, where the ratio $k/n$ is the code rate has the same significance as for block codes. The integer $K$ is a parameter known as the constraint length, it represent the number of $k$-tuple stages in the encoding shift register. The encoder transforms each sequence $\textbf{u}$ into a codeword sequence $\textbf{v}=G(\textbf{u})$. The key feature of convolution code is that a given $k$-tuple within $\textbf{u}$ does not uniquely define its associated $n$-tuple within $\textbf{v}$, since the encoding of each $k$-tuple is only a function of that $k$-tuple but is also a function of $K-1$ input $k$-tuples that precede it.
\begin{figure}[bpht!]
\includegraphics[scale=0.5]{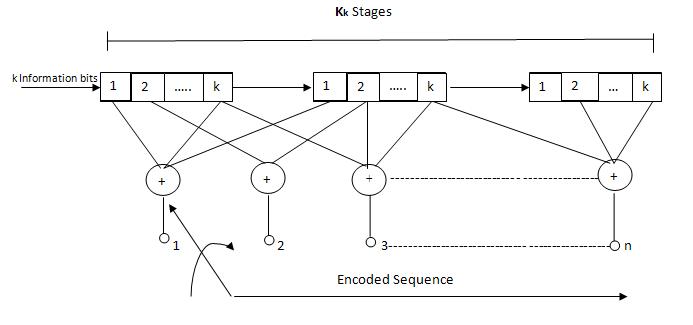}
\caption{Convolutional encoder with constraint length $K$ and rate $k/n$ \cite{17}.}
\end{figure}
A general convolutional encoder, is shown in Fig.1, is mechanized with $Kk$-stage shift register and $n$ modulo-2 adders, where $K$ is the constraint length.
\subsection{Golay Codes}
The binary form of the Golay code is one of the most important types of linear binary block codes. It is of particular significance since it is one of only a few examples of a nontrivial perfect code \cite{18}. A \emph{t-error-correcting} code can correct a maximum of \emph{t errors}. A perfect \emph{t-error correcting} code has the property that
every word lies within a distance of $t$ to exactly one code word. Equivalently, the code has $d_{min} = 2t + 1$, and covering radius $t$, where the covering radius $r$ is the smallest number such that every word lies within a distance of $r$ to a codeword.

\begin{theorem}
If there is an $(n,k)$ code with an alphabet of $q$ elements, and $d_{min}=2t+1$, then \\
$q^n\geqslant q^k\sum_{i=0}^t \binom{n}{i}(q-1)^i$
\end{theorem}
The inequality in Theorem 2 is known as the Hamming bound. Clearly, a code is perfect precisely when it attains equality in the Hamming bound. Two Golay codes do attain equality, making them perfect codes: the (23, 12) binary code with $d_{min} = 7$, and the $(11, 6)$ ternary code with $d_{min} = 5$. Both codes have the largest minimum distance
for any known code with the same values of $n$ and $k$. \\Golay was in search of perfect code when he noticed that $\binom{23}{0}+\binom{23}{1}+\binom{23}{2}+\binom{23}{3}=2^{11}=2^{23-12}$ which indicated the existence of a $(23,12)$ perfect code that could correct any combination of three or fewer random errors in a block of 23 bits. The $(23,12)$ Golay code can be generated using a method similar to CRC by using any of the following generator polynomial 
\begin{itemize}
\item $P_1(X)=X^{11}+X^{10}+X^6+X^5+X^4+X^2+1$ and
\item $P_2(X)=X^{11}+X^9+X^7+X^6+X^5+X+1$
\end{itemize}
this is due to the fact that \\$X^{23}+1=(X+1)P_1(X)P_2(X)$. For our analysis, we have constructed Golay codes using Hadamard matrix, which is explained below.
\subsubsection{Construction of extended binary Golay code $G_{24}$}
We have used the Hadamard matrix of Paley type of $p=11$ and $n=p+1=12$ for the construction of generator matrix for the Golay code. Hadamard Paley type is a normalized Hadamard matrix $H$ of order $n=p+1$ and $H$ is of the form
\begin{equation}
H=
\left[
\begin{array}{rr}
1&1\\
1&Q-I
\end{array}
\right]
\end{equation} 
where $I$ is a $p\times p$ identity matrix and $Q$ is a Jacobsthal matrix. Jacobsthal matrix $Q=(q_{ij})$ is a $p\times p$ matrix whose columns and rows are labeled as $0,1,2,\dots,p-1$ and $q_{ij}=\chi(j-i)$.
The Legendre symbol $\chi(i)$ is defined as:
\begin{displaymath}
\chi(i)=\left \{\begin{array}{ll}
0& \textrm{if $i$ is multiple of $p$}\\
1& \textrm{if the $\mbox{rem}(p \mid i)$ is a quadratic residue}\\
-1& \textrm{if the remainder is nonresidue}
\end{array}\right.
\end{displaymath}

For $p=11$,the Jacobsthal matrix $Q$ is as follows:
\begin{equation}
Q=
\left[
\begin{array}{rrrrrrrrrrr}
0&1&-&1&1&1&-&-&-&1&-\\
-&0&1&-&1&1&1&-&-&-&1\\
1&-&0&1&-&1&1&1&-&-&-\\
-&1&-&0&1&-&1&1&1&-&-\\
-&-&-&1&-&0&1&1&1&1&-\\
-&-&-&1&-&0&1&-&1&1&1\\
1&-&-&-&1&-&0&1&-&1&1\\
1&1&-&-&-&1&-&0&1&-&1\\
1&1&1&-&-&-&1&-&0&1&-\\
-&1&1&1&-&-&-&1&-&0&1\\
1&-&1&1&1&-&-&-&1&-&0
\end{array}
\right].
\end{equation}

Let $A_{p}=Q_{p}-I_{p}$, where $I$ is an identity matrix and $Q$ Jacobsthal matrix, then Hadamard of Paley type of order $n$ will be of the following from:
\begin{equation}
H_{p+1}=
\left[
\begin{array}{rrrr}
1&1&\ldots &1\\
1&&&\\
1&&&\\
\vdots&&A_p&\\
1&&&
\end{array}
\right].
\end{equation}\\
Without losing the generality, $A_{p+1}$ can be obtained similar to $H_{p+1}$. Now defining a code $C_{24}\subseteq V=F_2^{24}$ has $n=24$ and dimension $k=12$. Thus, its generator matrix $G_{24}$ has the form $G_{24}=(I_{12},A_{11+1})$, where $I_{12}$ is $12\times 12$ identity matrix and $A_{12}$ is a Hadamard matrix of order 12 of Paley type. Binary Golay code $C_{23}[23,12,7]$ is obtained by puncturing any column of $C_{24}[24,12,8]$.
\subsection{Reed-Muller Codes}
Reed-Muller codes are among the oldest and well known codes \cite{19}. Reed-Muller codes have many interesting properties. They form an infinite family of codes, and larger Reed-Muller codes can be constructed from
smaller ones. This particular observation lead us to show that Reed-Muller codes can be defined recursively. Assuming that we are given a vector space $\mathbb{F}_2^{2^m}$ and considering the ring $\mathcal{R}_m=\mathbb{F}_2[x_0,x_1,\cdots,x_m]$.
\begin{definition}
A \textit{Boolean monomial} is an element $p\in \mathcal{R}_m$ of the form:\\ 
$p=x_0^{r_0}x_1^{r_1}\cdots x_{m-1}^{r_{m-1}}$\\
where $r_i\in \mathbb{N}$ and $i\in \mathbb{Z}_m$. 
\end{definition}
A Boolean polynomial is a linear combination of Boolean monomials.

\begin{definition}
Given a Boolean monomial $p\in \mathcal{R}_m$, we say that $p$ is in reduced form if it is square free.
\end{definition}
For any Boolean monomial $q\in\mathcal{R}_m$, the reduced form $q$ is found by applying the following:

\begin{eqnarray*}
x_ix_j&=&x_jx_i \quad \mbox{as $\mathcal{R}_m$ is a commutative ring} \\ 
x_j^2&=&x_i \quad \mbox{as $0*0=0$ and $1*1=1$} .
\end{eqnarray*}
A Boolean polynomial in reduced form is simply a linear combination of reduced-form Boolean monomials (with coefficients in $\mathbb{F}_2$). \\
Consider the mapping $\psi:\mathcal{R}_m\to \mathbb{F}_2^{2^m}$, defined as follows:
\begin{eqnarray*}
\psi(0)&=& \underbrace{00\cdots 0}_{2^m}\\
\psi(1)&=& \underbrace{11\cdots 1}_{2^m}\\
\psi(x_0)&=& \underbrace{11\cdots 1}_{2^{m-1}}\underbrace{00\cdots 0}_{2^{m-1}}\\
\psi(x_1)&=&\underbrace{11\cdots 1}_{2^{m-2}}\underbrace{00\cdots 0}_{2^{m-2}}\underbrace{11\cdots 1}_{2^{m-2}}\underbrace{00\cdots 0}_{2^{m-2}} \\
\vdots & & \quad \vdots \\
\psi(x_i)&=&\underbrace{11\cdots 1}_{2^{m-i}}\underbrace{00\cdots 0}_{2^{m-i}}\cdots \\
\end{eqnarray*}
For any monomial $p\in\mathcal{R}_m$, to calculate $\psi(p)$, first we find its reduced form\\
$p'=x_{i_{1}}x_{i_{2}}\ldots x_{i_{r}}$, then $\psi(p)=\psi(x_{i_{1}})*\psi(x_{i_{2}})*\cdots *\psi(x_{i_{r}})$.\\
For the Reed-Muller code $\mathcal{RM}(r,m)$, the generator matrix is defined as follows:
\begin{equation}
G_{\mathcal{RM}(r,m)}=
\left[
	\begin{array}{c}
	\psi(1)\\
	\psi(x_0)\\
	\psi(x_1)\\
	\vdots \\
	\psi(x_{m-1})\\
	\psi(x_0x_1)\\
	\psi(x_0x-2)\\
	\vdots\\
	\psi(x_{m-2}x_{m-1})\\
	\psi(x_0x_1x_2)\\
	\vdots\\
	\psi(x_{m-r}x_{m-r+1}\ldots x_{m-1})
	
	\end{array}
\right].
\end{equation}
The matrix $G_{\mathcal{RM}(r,m)}$ has dimension $k\times n$, where $k=\sum_{i=0}^{r}\binom{m}{i}$ and $n=2^m$.

\section{Simulation Results and Discussion} A simplified diagram of OFDM Transmitter and Receiver used for our study is shown in Fig.2. The presented simulations has been performed for IEEE 802.11a standard (i.e N=64) modulated with 16 QAM. The statistics of peak and instantaneous power of OFDM symbols generated with a text as an input has been investigated with different channel coding techniques.
\begin{figure}[bpht!]
\centering
\includegraphics[scale=0.5]{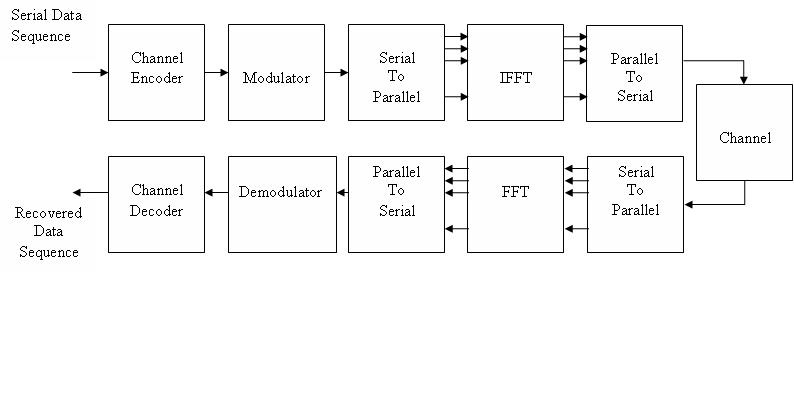} 
\caption{OFDM System.}
\label {fig:TxRx}
\end{figure}

The complementary cumulative distribution function (CCDF) has been used here to assess the PAPR reduction performance of the coded OFDM using various channel coding techniques against uncoded OFDM. The CCDF of the PAPR denotes the probability that the PAPR of the data block exceeds a given threshold $\xi$. The CCDF is given by (3). In all the simulations, value of PAPR has been studied for $CCDF=10^{-2}$.
Result of Hamming coded OFDM simulation is tabulated in Table.1 and CCDF vs PAPR plot is shown in Fig.3. From Fig.3, it is observed that Hamming Codes helps in reducing PAPR and maximum PAPR reduction has been observed for $m=6$.
\begin{figure}[bpht!]
\centering
\includegraphics[scale=0.8]{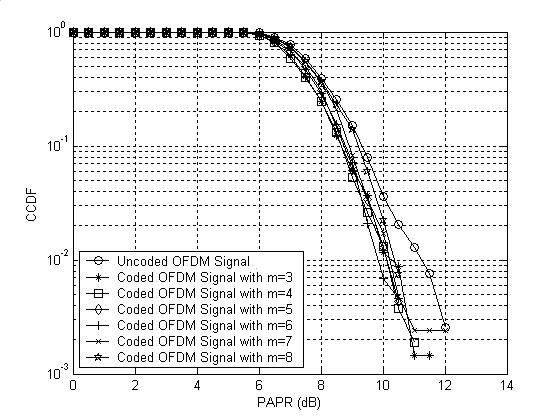} 
\caption{CCDF vs. PAPR curve of OFDM signal using Linear Block Code (Hamming) with different number of parity bits.}
\label {fig:Fig2}
\end{figure}

\begin{table*}[bpht!]
\begin{small}
\caption{Performance of Hamming coded OFDM signal for different values of parity bits (m).}
	\centering
		\begin{tabular}{|p{3cm}|p{3cm}|p{2.5cm}|p{2.5cm}|p{1.5cm}|}
		\hline
		Hamming Code Configuration&PAPR of Uncoded OFDM Signal (dB)&PAPR of Coded OFDM Signal (dB)& Reduction in PAPR (dB) & Code Rate $(k/n)$\\
		\hline
		Parity bit (m)=3&11.2742&10.3065&0.9677&0.5714\\
		\hline
		Parity bit (m)=4&11.2742&10.1452&1.1290&0.7333\\
		\hline
		Parity bit (m)=5&11.2742&10.1452&1.1290&0.8484\\
		\hline
		Parity bit(m)=6&11.2742&09.8548&1.4194&0.9047\\
		\hline
		Parity bit (m)=7&11.2742&10.2419&1.0323&0.9448\\
		\hline
		Parity bit (m)=8&11.2742&10.4032&0.8710&0.9686\\
		\hline
		\end{tabular}
\end{small}
\end{table*}

From Fig.4 and Table.2, it is observed that PAPR reduction performance of cyclic code is slightly better than the Hamming code and the maximum PAPR reduction has been obtained for $m=4$.
\begin{figure}[bpht!]
\centering
\includegraphics[scale=0.8]{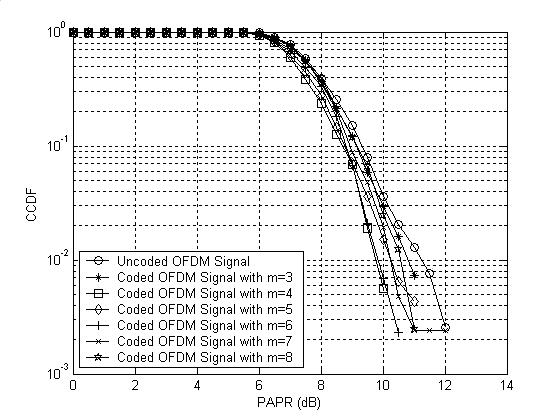} 
\caption{CCDF vs. PAPR curve of OFDM signal using cyclic codes with different number of parity bits.}
\label {fig:CyclicCodeResults}
\end{figure}
\begin{table*}[bpht!]
\begin{small}
	\centering
	\caption{Performance of Cyclic coded OFDM signal for different values of parity bits (m).}
		\begin{tabular}{|p{3cm}|p{3cm}|p{2.5cm}|p{2.5cm}|p{1.5cm}|}
		\hline
		Cyclic Code Configuration&PAPR of Uncoded OFDM Signal (dB)&PAPR of Coded OFDM Signal (dB)& Reduction in PAPR (dB) & Code Rate $(k/n)$\\
		\hline
		Parity bit (m)=3&11.2742&10.8226&0.4516&0.5714\\
		\hline
		Parity bit (m)=4&11.2742&9.7581&1.5161&0.7333\\
		\hline
		Parity bit (m)=5&11.2742&10.1452&1.1290&0.8484\\
		\hline
		Parity bit (m)=6&11.2742&09.8548&1.4194&0.9047\\
		\hline
		Parity bit (m)=7&11.2742&10.2419&1.0323&0.9448\\
		\hline
		Parity bit (m)=8&11.2742&10.4032&0.8710&0.9686\\
		\hline
		\end{tabular}
\end{small}
\end{table*}

The connection vectors or polynomial generators of a convolutional code are usually selected based on the code's free distance property \cite{17}. The first criterion is to select a code that does not have catastrophic error propagation and that has the maximum free distance for the given rate and constraint length. Then the number of paths at the free distance or the number of data bit errors the path represents should be minimized. For our investigation, we have referred a list of the best known code of rate $\frac{1}{2}$, K=3 to 14, and rate 1/3, K=3 to 14, which was compiled by Odenwalder \cite{20} based on the above criteria. Generator vectors of the convolution codes are listed in octal form in Table.3 \cite{17}. 
\begin{table*}[bpht!]
\begin{center}
\begin{small}
\caption{Convolutional Codes Generator Vectors.}
\begin{tabular}{|c|c|p{1.25in}|c|c|p{1.8in}|}
\hline
\textbf{Code Rate}& \textbf{K}& \textbf{Generators in Octal Form}& \textbf{Code Rate}& \textbf{K}& \textbf{Generators  in Octal Form}\\
\hline
1/2&3&[5 7]&1/3&3&[5 7 7] \\	\hline 
1/2&4&[15 17]&1/3&4&[13 15 17] \\	\hline 
1/2&5&[23 35]&1/3&5&[25 33 37]	\\	\hline 
1/2&6&[53 75]&1/3&6&[47 53 75]	\\	\hline 
1/2&7&[133 171]&1/3&7&[133 145 175]	\\	\hline 
1/2&8&[247 371]&1/3&8&[225 331 367]	\\	\hline 
1/2&9&[561 753]&1/3&9&[557 663 711] \\ \hline 
1/2&10&[1,167 1,545]&1/3&10&[1,117 1,365 1,633]\\ \hline
1/2&11&[2,335 3,661]&1/3&11&[2,353 2,671 3,175]\\ \hline
1/2&12&[4,335 5,723]&1/3&12&[4,767 5,723 6,265]\\ \hline
1/2&13&[10,533 17,661]&1/3&13&[10,533 10,675 17,661]\\ \hline
1/2&14&[21,675 27,123]&1/3&14&[21,645 35,661 37,133]\\ \hline
\end{tabular}
\end{small}
\end{center}
\end{table*}

Fig.5 and 6 shows CCDF vs. PAPR curve of direct OFDM signal along with CCDF vs. PAPR curve for convolutionally coded OFDM signal with code rate 1/2 and different values of constraint length K. From Fig.5, it is observed that the PAPR reduction with convolution code is further better than the cyclic codes. Here the performance of convolution code with rate 1/2 and of different constraint length, from K=3 to 14 has been investigated. It has been observed that there is reduction of PAPR by more than $1.5$ dB for all the value of $K$ and the maximum reduction is observed for $K=6$. The results of simulation has been tabulated in Table.4.

\begin{figure}[bpht!]
\centering
\includegraphics[scale=0.8]{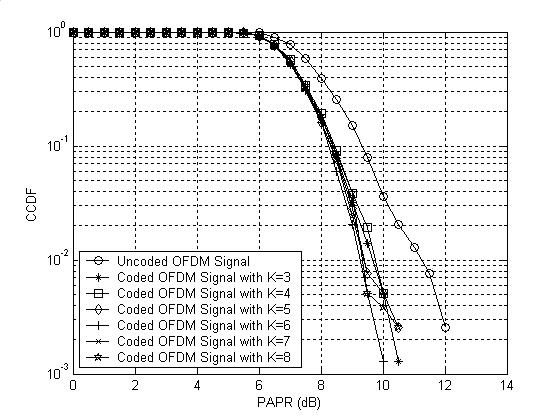} 
\caption{CCDF vs. PAPR curve of OFDM signal using convolution codes with code rate $\frac{1}{2}$ and constraint length $(K=3-8)$.}
\end{figure}

\begin{figure}[bpht!]
\centering
\includegraphics[scale=0.8]{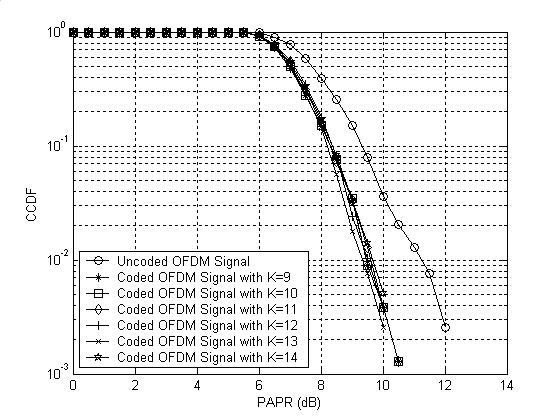} 
\caption{CCDF vs. PAPR curve of OFDM signal using convolution codes with code rate $\frac{1}{2}$ and constraint length $(K=9-14)$.}
\label {fig:Fig5}
\end{figure}
\clearpage
\begin{table*}[bpht!]
\begin{small}
	\centering
	\caption{Performance of rate 1/2 convolution coded OFDM signal for different values of constraint length(K).}
		\begin{tabular}{|p{3cm}|p{3cm}|p{2.5cm}|p{2.5cm}|}
		\hline
		Convolution Code Constraint Length &PAPR of Uncoded OFDM Signal (dB)&PAPR of Coded OFDM Signal (dB)& Reduction in PAPR (dB) \\
		\hline
		K=3&11.2742&9.6935&1.5807\\ \hline
		K=4&11.2742&9.7581&1.5161\\ \hline
		K=5&11.2742&9.4032&1.8710\\ \hline
		K=6&11.2742&9.2742&2.0000\\ \hline
		K=7&11.2742&9.4355&1.8387\\ \hline
		K=8&11.2742&9.3710&1.9082\\ \hline
		K=9&11.2742&9.6613&1.6129\\ \hline
		K=10&11.2742&9.4677&1.8065\\ \hline
		K=11&11.2742&9.6290&1.6452\\ \hline
		K=12&11.2742&9.5323&1.7419\\ \hline
		K=13&11.2742&9.3710&1.9032\\ \hline
		K=14&11.2742&9.6935&1.5807\\ \hline
		\end{tabular}
\end{small}
\end{table*}


\begin{figure}[bpht!]
\centering
\includegraphics[scale=0.8]{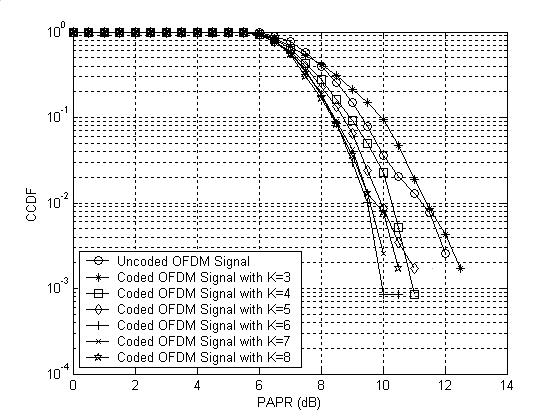} 
\caption{CCDF vs. PAPR curve of OFDM signal using convolution codes with code rate $\frac{1}{3}$ and constraint length $(K=3-8)$.}
\end{figure}

\begin{figure}[bpht!]
\centering
\includegraphics[scale=0.8]{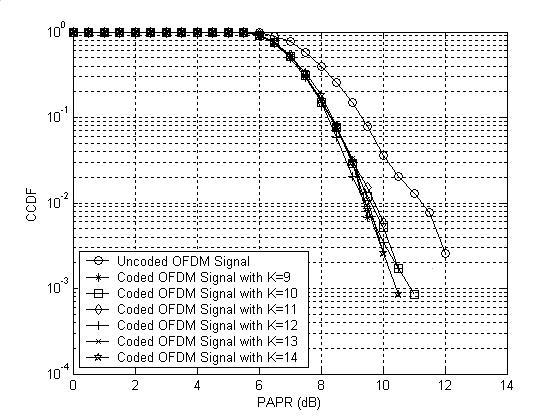} 
\caption{CCDF vs. PAPR curve of OFDM signal using convolution codes with code rate $\frac{1}{3}$ and constraint length (K=9-14).}
\end{figure}
Fig.7 and 8 shows CCDF versus PAPR curve of direct OFDM signal along with CCDF vs. PAPR curve for convolutionally coded OFDM signal with code rate 1/3 and different values of constraint length K. The performance of convolution code with rate 1/3 and of different constraint length, from K=3 to 14 has been investigated. It has been observed that there is reduction of PAPR by more than $1$ dB for all the value of $K$ and the maximum reduction is observed for $K=9$. The results of simulation has been tabulated in Table.5. Furthermore from Table.4 and Table.5, it can be inferred that the PAPR reduction performance of convolution code with rate 1/2 is slightly better and less complex from implementation perspective than the convolution code with rate 1/3.

\begin{table*}[bpht!]
\begin{small}
	\centering
	\caption{Performance of rate 1/3 convolution coded OFDM signal for different values of constraint length(K).}
		\begin{tabular}{|p{3cm}|p{3cm}|p{2.5cm}|p{2.5cm}|}
		\hline
		Convolution Code Constraint Length &PAPR of Uncoded OFDM Signal (dB)&PAPR of Coded OFDM Signal (dB)& Reduction in PAPR (dB) \\
		\hline
		K=3&11.2742&11.4355&-0.1613\\ \hline
		K=4&11.2742&10.3065&0.9677\\ \hline
		K=5&11.2742&9.9516&1.3226\\ \hline
		K=6&11.2742&9.500&1.7742\\ \hline
		K=7&11.2742&9.5968&1.6774\\ \hline
		K=8&11.2742&9.8226&1.4516\\ \hline
		K=9&11.2742&9.4032&1.8710\\ \hline
		K=10&11.2742&9.5968&1.6774\\ \hline
		K=11&11.2742&9.7258&1.5484\\ \hline
		K=12&11.2742&9.4032&1.8710\\ \hline
		K=13&11.2742&9.5323&1.7419\\ \hline
		K=14&11.2742&9.5000&1.7742\\ \hline
		\end{tabular}
\end{small}
\end{table*}
Fig.9 shows the CCDF vs PAPR curve of [23,12,7] and [24,12,8] Golay coded OFDM signal. From Fig.9, it can be easily inferred that the PAPR reduction performance of [23,12,7] Golay coded OFDM is better than the [24,12,8] Golay coded OFDM signal and the PAPR reduction is approximately 2 dB.

\begin{figure}[bpht!]
\includegraphics[scale=0.8]{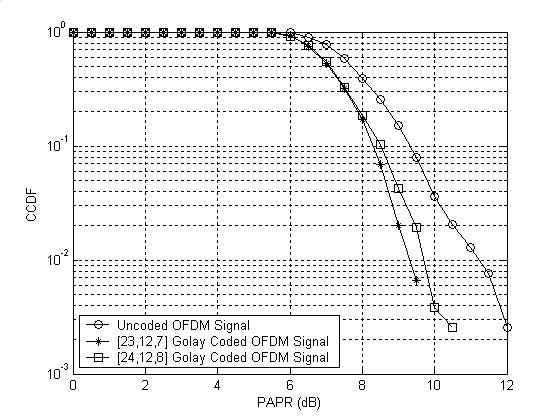}
\caption{CCDF vs. PAPR curve of OFDM signal using Golay codes.}
\end{figure}

\begin{figure}[bpht!]
\centering
\includegraphics[scale=0.8]{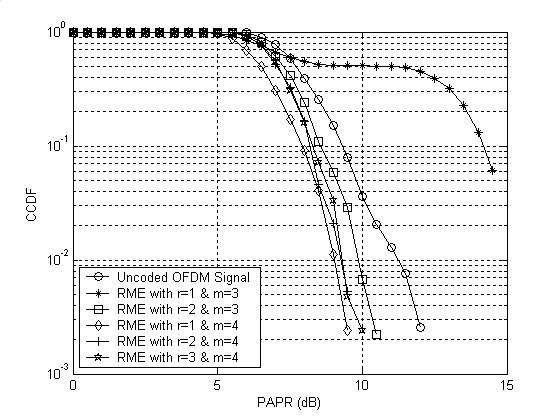}
\caption{CCDF vs. PAPR curve of OFDM signal using Reed-Muller code with code different values of $r$ and $m$.}
\end{figure}

Fig.10 and 11 shows the CCDF vs PAPR curve for Reed-Muller coded OFDM signal with different values of $r$ and $m$ and the result of simulation has been tabulated in Table.6. From Table.6, it is observed that the maximum PAPR reduction obtained is equal to 2.2362 dB for $r=1$ and $m=4$. The PAPR reduction performance of Reed-Muller code is better than the Golay code as evident from Figs.9,10 and 11.
\begin{figure}[bpht!]
\centering
\includegraphics[scale=0.8]{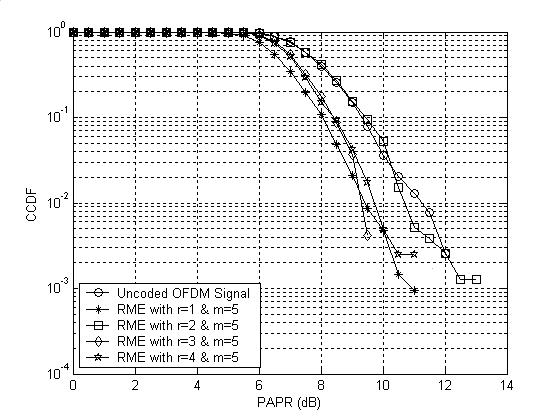}
\caption{CCDF vs. PAPR curve of OFDM signal using Reed-Muller code with code different values of $r$ and $m$.}
\end{figure}

\begin{table*}[bpht!]
\begin{small}
	\centering
	\caption{Performance of Reed-Muller coded OFDM signal for different values of $r$ and $m$.}
		\begin{tabular}{|p{3cm}|p{3cm}|p{2.5cm}|p{2.5cm}|p{1.5cm}|}
		\hline
		Reed-Muller code figuration&PAPR of Uncoded OFDM Signal (dB)&PAPR of Coded OFDM Signal (dB)& Reduction in PAPR (dB) & Code Rate $(k/n)$\\		\hline
		$r=1$ and $m=3$&11.2742&14.6717&-&0.5000\\ \hline
		$r=2$ and $m=3$&11.2742&9.8675&1.4067&0.8750\\ \hline
		$r=1$ and $m=4$&11.2742&9.0380&2.2362&0.3125\\ \hline
		$r=2$ and $m=4$&11.2742&9.2800&1.9942&0.6875\\ \hline
		$r=3$ and $m=4$&11.2742&9.3145&1.9597&0.9375\\ \hline
		$r=1$ and $m=5$&11.2742&9.4355&1.8387&0.1875\\ \hline
		$r=2$ and $m=5$&11.2742&10.7258&0.5484&0.5000\\ \hline
		$r=3$ and $m=5$&11.2742&9.3065&1.9677&0.8125\\ \hline
		$r=4$ and $m=5$&11.2742&9.7258&1.5485&0.9687\\ \hline
		
		\end{tabular}
\end{small}
\end{table*}

Fig.12 shows the PAPR reduction performance of various channel coding techniques considered namely,Hamming code, cyclic code, convolutional code with rate 1/2 and 1/3 respectively,Golay code and Reed-Muller code. The PAPR reduction performance of all the considered channel coding techniques has been tabulated in Table.7. From Table.7 it can be easily inferred that the PAPR reduction performance of Reed-Muller code is better than the other considered codes.
\begin{figure}[bpht!] 
\centering
\includegraphics[scale=0.8]{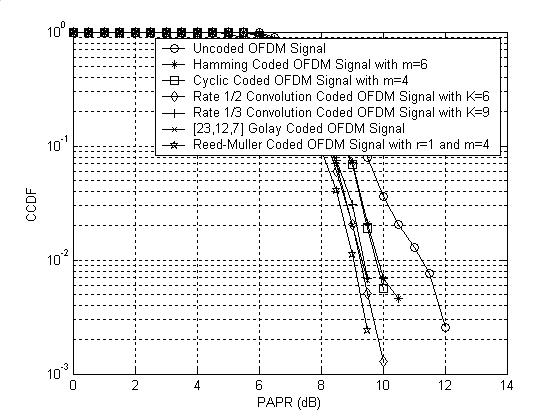}
\caption{CCDF vs. PAPR curve of coded OFDM signal using different channel coding techniques.}
\end{figure}

\begin{table*}[bpht!]
\begin{small}
	\centering
	\caption{Performance of error control coded OFDM signal for different channel codes.}
		\begin{tabular}{|p{3cm}|p{3cm}|p{2.5cm}|p{2.5cm}|p{1.5cm}|}
		\hline
		Channel Coding Technique&PAPR of Uncoded OFDM Signal (dB)&PAPR of Coded OFDM Signal (dB)& Reduction in PAPR (dB) & Code Rate $(k/n)$\\		\hline
		Hamming code with m=6&11.2742&9.8548&1.4194&0.9047\\ \hline
		Cyclic code with m=4&11.2742&9.7581&1.5161&0.7333\\ \hline
		Convolution code with K=6 and R=1/2&11.2742&9.2742&2.0000&0.5000\\ \hline
		Convolution code with K=9 and R=1/3&11.2742&9.4032&1.8710&0.3333\\ \hline
		[23,12,7] Golay code&11.2742&9.3065&1.9677&0.5217 \\ \hline
		Reed-Muller code with r=1 and m=4&11.2742&9.0380&2.2362&0.3125\\ \hline
		\end{tabular}
\end{small}
\end{table*}

From Table.7, it is clear that the use of channel coding for OFDM signal reduces the aperiodic autocorrelation between OFDM symbols on subcarriers as a result of which PAPR of the coded OFDM signal reduces though to a very small extent at most by 2 dB. This fact is the direct result of equation(5). The channel coding techniques can be arranged in the following order based on their PAPR reduction performance:
\begin{enumerate}
\item Reed-Muller code with $r=1$ and $m=4$
\item Convolution code, with rate 1/2 and K=6
\item $[23,12,7]$ Golay code
\item Convolution code, with rate 1/3 and K=9
\item Cyclic code with m=4
\item Hamming code with m=6.
\end{enumerate}

\section{Conclusion}
OFDM is one of the promising multicarrier modulation technique which offer very high data rates over the wireless medium without getting affected by ISI and fading over the wireless medium.
\par
In this paper, we have analyzed the PAPR of BPSK modulated OFDM signal. As an outcome of the analysis, it is found that the PAPR is completely characterized by the aperiodic autocorrelation of OFDM symbols on subcarriers. 
\par To reduce the aperiodic autocorrelation of OFDM symbol on subcarriers, we have used various channel coding techniques; especially linear block codes such as Hamming code, cyclic code, convolution code, Golay code and Reed-Muller codes, for the OFDM signal. As a result of this we have compared the PAPR reduction performance of the considered channel codes and found that the PAPR reduction performance of Reed-Muller code with $r=1$ and $m=4$ is the best among the considered channel codes. Moreover, from the perspective of hardware complexity of encoder and decoder, PAPR reduction performance of convolution code is found to be optimum as compared to other channel coding techniques. Though the reduction in PAPR achieved is at most 2 dB, one of the major advantages of the discussed study is that it can be combined with selective mapping (SLM) or other conventional PAPR reduction techniques for further reduction in PAPR.
\bibliographystyle{IEEEtran}
\bibliography{ref}

\begin{thebibliography}{10}
\providecommand{\url}[1]{#1}
\csname url@samestyle\endcsname
\providecommand{\newblock}{\relax}
\providecommand{\bibinfo}[2]{#2}
\providecommand{\BIBentrySTDinterwordspacing}{\spaceskip=0pt\relax}
\providecommand{\BIBentryALTinterwordstretchfactor}{4}
\providecommand{\BIBentryALTinterwordspacing}{\spaceskip=\fontdimen2\font plus
\BIBentryALTinterwordstretchfactor\fontdimen3\font minus
  \fontdimen4\font\relax}
\providecommand{\BIBforeignlanguage}[2]{{%
\expandafter\ifx\csname l@#1\endcsname\relax
\typeout{** WARNING: IEEEtran.bst: No hyphenation pattern has been}%
\typeout{** loaded for the language `#1'. Using the pattern for}%
\typeout{** the default language instead.}%
\else
\language=\csname l@#1\endcsname
\fi
#2}}
\providecommand{\BIBdecl}{\relax}
\BIBdecl

\bibitem{1}
M.~Nakhai, ``Multicarrier transmission,'' \emph{Signal Processing, IET},
  vol.~2, no.~1, pp. 1 --14, march 2008.

\bibitem{2}
H.~Gacanin and F.~Adachi, ``A comprehensive performance comparison of ofdm/tdm
  using mmse-fde and conventional ofdm,'' in \emph{Vehicular Technology
  Conference, 2008. VTC Spring 2008. IEEE}, may 2008, pp. 1404 --1408.

\bibitem{3}
, ``Broadband radio access networks (bran);hiperlan type 2;data link control
  (dlc) layer part 1: Basic data transport functions,'' [Available Online]
  http://www.etsi.org/deliver/etsi\_ts/101700\_101799/10176101/01.01.01\_60/ts\_10176101v010101p.pdf,
  2000.

\bibitem{4}
IEEE, ``Draft supplement to standard for telecommunications and information
  exchange between systems-lan/man specific requirements-part 11: Wireless mac
  and phy specifications: High speed physical layer in the 5 ghz band,
  ieee802.11,'' IEEE, Tech. Rep. P802.11a/D6.0(1999), May 1999.

\bibitem{5}
H.~Ochiai and H.~Imai, ``Performance of block codes with peak power reduction
  for indoor multicarrier systems,'' in \emph{Vehicular Technology Conference,
  1998. VTC 98. 48th IEEE}, vol.~1, may 1998, pp. 338 --342 vol.1.

\bibitem{6}
X.~Li and J.~Cimini, L.J., ``Effects of clipping and filtering on the
  performance of ofdm,'' \emph{Communications Letters, IEEE}, vol.~2, no.~5,
  pp. 131 --133, may 1998.

\bibitem{7}
A.~Jones, T.~Wilkinson, and S.~Barton, ``Block coding scheme for reduction of
  peak to mean envelope power ratio of multicarrier transmission schemes,''
  \emph{Electronics Letters}, vol.~30, no.~25, pp. 2098 --2099, dec 1994.

\bibitem{8}
T.~Wilkinson and A.~Jones, ``Minimisation of the peak to mean envelope power
  ratio of multicarrier transmission schemes by block coding,'' in
  \emph{Vehicular Technology Conference, 1995 IEEE 45th}, vol.~2, jul 1995, pp.
  825 --829 vol.2.

\bibitem{9}
C.-Y. Chen, C.-H. Wang, and C.~chao Chao, ``Convolutional codes for
  peak-to-average power ratio reduction in ofdm,'' in \emph{Information Theory,
  2003. Proceedings. IEEE International Symposium on}, june-4 july 2003, p.~5.

\bibitem{10}
J.~Davis and J.~Jedwab, ``Peak-to-mean power control in ofdm, golay
  complementary sequences, and reed-muller codes,'' \emph{Information Theory,
  IEEE Transactions on}, vol.~45, no.~7, pp. 2397 --2417, nov 1999.

\bibitem{11}
K.~Paterson and V.~Tarokh, ``On the existence and construction of good codes
  with low peak-to-average power ratios,'' in \emph{Information Theory, 2000.
  Proceedings. IEEE International Symposium on}, 2000, p. 217.

\bibitem{12}
J.~Tellado, \emph{Multicarrier Modulation with Low PAR: Applications to DSL and
  Wireless}.\hskip 1em plus 0.5em minus 0.4em\relax Boston: Kluwer Academic
  Publisher, 2000.

\bibitem{13}
R.~Nee and R.Prasad, \emph{OFDM for Wireless Multimedia Communications}.\hskip
  1em plus 0.5em minus 0.4em\relax Artech House, 2000.

\bibitem{14}
S.~Narahashi and T.~Nojima, ``New phasing scheme of n-multiple carriers for
  reducing peak-to-average power ratio,'' \emph{Electronics Letters}, vol.~30,
  no.~17, pp. 1382 --1383, aug 1994.

\bibitem{15}
M.~K. Sanjay~Singh and H.~Mruthyunjaya, ``Reduction of peak-to-average power
  ratio of an ofdm signal using scrambling coding,'' \emph{International
  Journal of Computer Science and Communication Technology}, vol.~1, no.~2, pp.
  109--115, 2009.

\bibitem{16}
S.~Lin and D.~J. Jr., \emph{Error Control Coding: Fundamentals and
  Applications}.\hskip 1em plus 0.5em minus 0.4em\relax Upper Saddle River, New
  Jersey: Prentice Hall, 2004.

\bibitem{17}
J.~Proakis, \emph{Digital Communications}, 4th~ed.\hskip 1em plus 0.5em minus
  0.4em\relax New York: McGraw-Hill, 2001.

\bibitem{18}
\BIBentryALTinterwordspacing
M.~T.~E. Golay, ``Multi-slit spectrometry,'' \emph{J. Opt. Soc. Am.}, vol.~39,
  no.~6, pp. 437--437, Jun 1949. [Online]. Available:
  \url{http://www.opticsinfobase.org/abstract.cfm?URI=josa-39-6-437}
\BIBentrySTDinterwordspacing

\bibitem{19}
F.~J. MacWilliams and N.~J.~A. Sloane, \emph{The Theory of Error Correcting
  Codes}, 1st~ed.\hskip 1em plus 0.5em minus 0.4em\relax Amsterdam: Elsevier
  Science, 1977.

\bibitem{20}
J.~P. Odenwalder, \emph{Error Control Coding Handbook}.\hskip 1em plus 0.5em
  minus 0.4em\relax San Diego, California: Linkabit Corporation, 1976.

\end{thebibliography}

\end{document}